\documentclass[acmsmall]{acmart}
\usepackage{rotating}
\AtBeginDocument{%
  \providecommand\BibTeX{{%
    \normalfont B\kern-0.5em{\scshape i\kern-0.25em b}\kern-0.8em\TeX}}}

\acmJournal{CSUR}

\begin{document}

\title{Serverless Computing: A Survey of Opportunities, Challenges, and Applications}

\author{Hossein Shafiei}
\email{shafiei@kntu.ac.ir}
\orcid{0000-0002-4125-3595}
\email{shafiei@kntu.ac.ir}
\affiliation{%
  \institution{K. N. Toosi University of Technology}
  \streetaddress{P.O. Box  15875-4416}
  \city{Tehran}
  \country{Iran}
  \postcode{43017-6221}
}

\author{Ahmad Khonsari}
\affiliation{%
  \institution{University of Tehran}
  \city{Tehran}
  \country{Iran}}
\email{ak@ipm.ir}

\author{Payam Mousavi}
\affiliation{%
  \institution{University of Tehran}
  \city{Tehran}
  \country{Iran}}
\email{pa.mousavi@ut.ac.ir}

\renewcommand{\shortauthors}{Shafiei, et al.}

\begin{abstract}
The emerging serverless computing paradigm has attracted attention from both academia and industry. This paradigm brings benefits such as less operational complexity, a pay-as-you-go pricing model, and an auto-scaling feature. The paradigm opens up new opportunities and challenges for cloud application developers. In this paper, we present a comprehensive overview of the past development as well as the recent advances in research areas related to serverless computing. First, we survey serverless applications introduced in the literature and summarize the challenges they have faced. We categorize applications in eight domains and separately discuss the objectives and the viability of the serverless paradigm in each of those domains. We then classify those challenges into nine topics and survey the proposed solutions for each of them. Finally, we present the areas that need further attention from the research community and identify open problems.
\end{abstract}

\ccsdesc[500]{Computer systems organization~Cloud computing}

\keywords{Cloud Services, Serverless Computing, Function-as-a-Service (FaaS)}

\maketitle

\section{Introduction}
Large technology companies such as Amazon, Google, and Microsoft offer serverless platforms under various brand names. Although the specifics of the services may differ the essential idea behind the offered services is almost the same i.e. by rendering computation to the pay-as-you-go model, serverless computing tries to achieve auto-scaling while providing affordable computation services \cite{jonas2019cloud}. 


Serverless computing differs from traditional cloud computing concepts (we refer to them as serverful in this paper) in the sense that the infrastructure and the platforms in which the services are running are hidden from customers. In this approach, the customers are only concerned with the desired functionality of their application and the rest is delegated to the service provider.

There are successful commercial implementations of this model. Amazon introduced Lambda \footnote{ \texttt{https://aws.amazon.com/lambda/}} in 2014 and later Google Cloud Functions \footnote{\texttt{https://cloud.google.com/functions/}}, Microsoft Azure Functions \footnote{\texttt{https://azure.microsoft.com/}} and IBM OpenWhisk \footnote{\texttt{https://openwhisk.apache.org/}} were launched in 2016. Since then, many studies have focused on the challenges and open problems of this concept. Some of the previous studies are skeptical about the potentials of serverless computing due to the poor performance of their case studies \cite{hellerstein2018serverless}. In contrast, others believe that serverless computing will become the face of cloud computing and the performance issues will be addressed eventually \cite{jonas2019cloud}.

The aim of the serverless services is threefold: (1) relieve the users of cloud services from dealing with the infrastructures or the platforms, (2) convert the billing model to the pay-as-you-go model, (3) auto-scale the service per customers' demand. As a result in a truly serverless application, the execution infrastructure is hidden from the customer and the customer only pays for the resources they actually use. The service is designed such that it can handle request surges rapidly by scaling automatically. The basic entities in serverless computing are functions. The customer registers their functions in the service provider. Then, those functions can be invoked either by an event or per users' request. The execution results are sent back to the customer. The invocation of the functions is delegated to one of the available computation nodes inside the service provider. Usually, these nodes are cloud containers such as Docker \cite{merkel2014docker} or an isolated runtime environment \cite{hendrickson2016serverless}.

Though the concept of serverless computing is relatively new it has paved its way into many real-world applications ranging from online collaboration tools to the Internet of Things (IoT). We survey the papers that introduce real-world serverless applications and categorize them into eight different domains. We summarize the objectives (as justified by the authors) for migrating to serverless services for each of the application domains. We further assess the aptness of the serverless paradigm for each application domain based on the arguments made by the authors, the obtained results, and the challenges they reported. Table \ref{tab:the_table} lists the application domains, migration objectives, and assessments.

There are several challenges that serverless services are currently facing. There exist some surveys and literature reviews that discuss those challenges \cite{baldini2017serverless, jonas2019cloud, buyya2018manifesto, castro2019server, wu2020survey}. Our approach is different, in that, instead of focusing on the inherent obstacles and shortcomings of the concept, we analyze the challenges reported in each of the surveyed application domains. Then, we categorize and discuss the existing solutions for those challenges. We bring out the areas that need further attention from the research community. We also discuss the limitations of those solutions and identify open problems. Some of the challenges surveyed in this paper are common between various application domains such as the topic of providing security and privacy in serverless services. And some of them are domain-specific, such as the issues of scheduling, pricing, caching, provider management, and function invocation.


In this paper, we also discuss the opportunities presented by serverless computing. We emphasize that serverless services are more customer-friendly as they relieve customers from the intricacies of deployment. They are also more affordable in some cloud computing scenarios that we will discuss later in this paper. We argue that new market places are emerging around these services, which implies new business opportunities.

The rest of this paper is organized as follows. Section 2 presents definitions and characteristics of serverless services. Section 3 focuses on the opportunities that the serverless computing model offers. Section 4 discusses the application domains. Section 5 surveys the challenges toward the vast adoption of the concept. Section 6 concludes the paper.

\section{Definition and Characteristics}

\begin{figure}[t]
\centering
\includegraphics[width=2.4in]{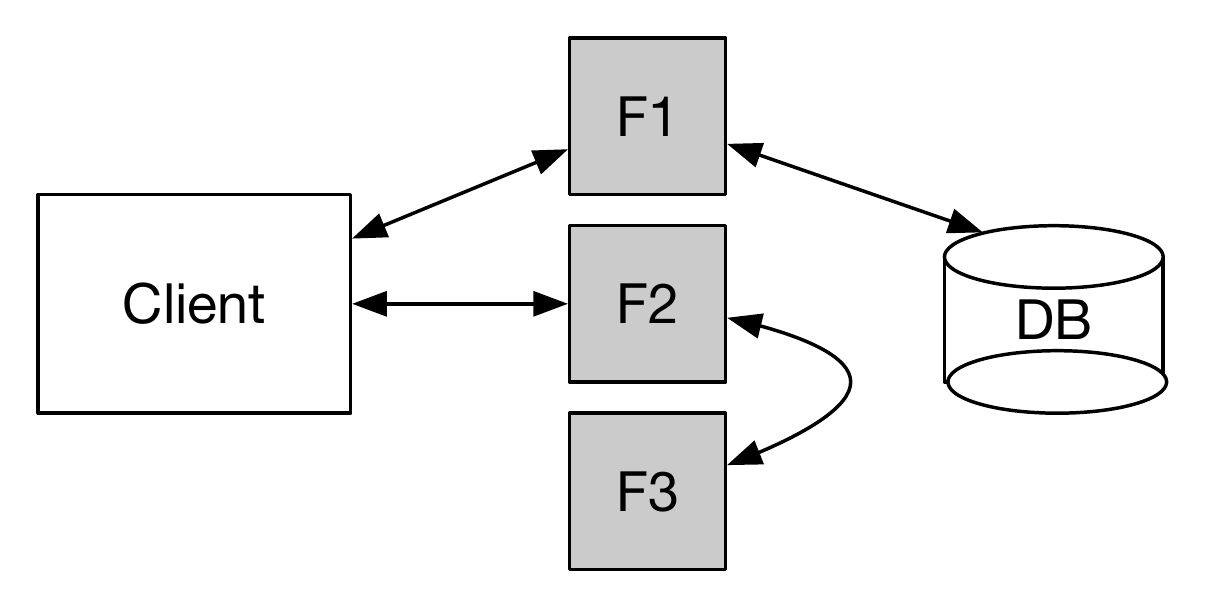}
\caption{An example of a serverless application.}
\label{fig:anatomy}
\end{figure}

\subsection{Definitions}
There is no formal definition for the concept of serverless computing and its various services. So, here we present the most commonly acknowledged definitions.
\subsubsection{FaaS}
Function as a service (FaaS) is a paradigm in which the customers are able to develop, run, and manage application functionalities without the trouble of building and maintaining the infrastructure. 
\subsubsection{BaaS}
Backend as a Service (BaaS) is an online service that handles a specific task over the cloud, such as authentication or notification. Both BaaS and FaaS require no resource management from the customers. While FaaS only offers to execute users' functions, BaaS offers a complete online service.

\subsubsection{Serverless service}
A serverless service can be viewed as a generalization of FaaS and BaaS that incorporates the following characteristics: 

\begin{enumerate}
\item The execution environment should be hidden from the customer i.e. the computation node, the virtual machine, the container, its operating system and etc. are all hidden from the customer.
\item The provider\footnote{Throughout this paper by a service provider we mean the organization (or company) that serves customers in a serverless fashion.} should provide an auto-scaling service i.e. the resources should be made available to the customer instantly per demand.
\item The billing mechanism should only reflect the number of resources the customer actually uses i.e. pay-as-you-go billing model.

\item The provider does its best effort to complete the customer's task as soon as it receives the request and the execution duration is bounded.

\item The basic elements in serverless services are functions. The functions are not hidden from the provider. The provider knows their dependencies to external libraries, run-time environments, and state during and after execution.
\end{enumerate}

A serverless application is usually comprised of two parts:
\begin{itemize}
\item A client \footnote{Note that, the client part is not essential in every serverless scenario i.e. often functions are directly called by other functions or they are triggered by an external event}. The client implements most of the application logic. It interacts with two sides i.e. the end-user and the provider, invoking functions on one side and translating the results into usable views for the other side.
\item Registered functions on a provider. Functions are uploaded to the provider. The provider invokes a copy of the function according to the user's request or based on a predefined event.
\end{itemize}

Figure \ref{fig:anatomy} depicts an example of a serverless application. In this example, the client invokes function F1 which retrieves data from the database, performs some operations on it, and sends the result back to the client. In another scenario, the client invokes F2, F2 invokes another function called F3, the result is sent back to F2 and then it sends its result to the client. In this scenario, we can see a sequence of function executions. Later in this paper, we argue that these sequences are important for performance optimization.

\begin{figure}[t]
\centering
\includegraphics[width=2.6in]{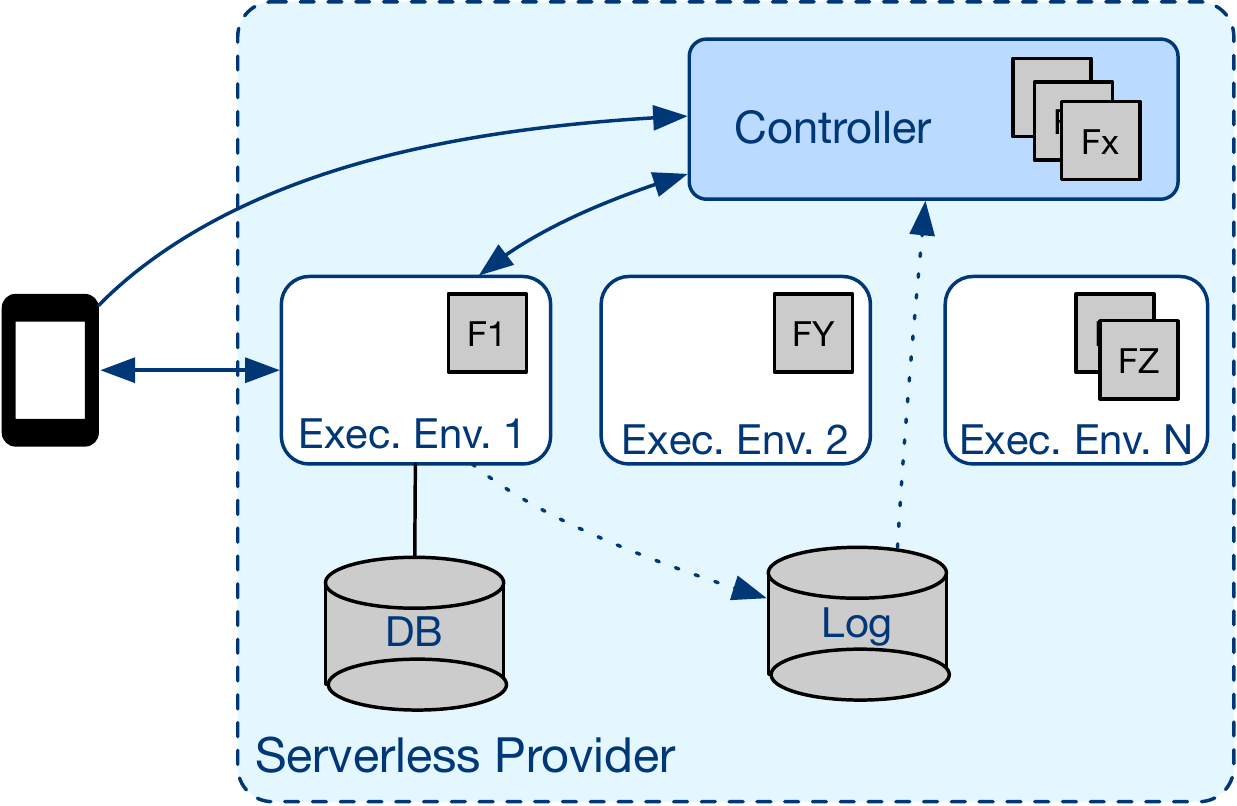}
\caption{An example of a invocation scenario in a serverless provider. Note that, this figure only depicts the concept and the actual implementation may differ}
\label{fig:provider}
\end{figure}

Providers accept customers' functions and store them. Upon receiving an invocation request, the provider assigns the function to one of its computation nodes for execution. The parameters upon which the provider selects the execution node have a great impact on the performance of the system which is thoroughly surveyed in Section \ref{sec:challenges} of this paper. The node (which can be a virtual machine, a container, or any sandbox execution environment) executes the function, sends back the result to the client, and sends the execution log to the provider. The provider can use these logs to improve further execution of the function. Figure \ref{fig:provider} shows an example of such a scenario.

\section{Opportunities}
In this section, we discuss the opportunities that serverless computing offers.  

\subsection{No deployment and maintenance complexity}
The very first and foremost opportunity that serverless computing offers is to relieve users from managing the infrastructure, which is now already accomplished in Infrastructure-as-a-Service (IaaS), however, users still have to manage their virtual resources i.e, installing and configuring related packages and libraries. Certainly, Platform-as-a-Service (PaaS) providers such as Heroku \cite{heroku} have made the management slightly easier, although, users still have to configure the application to match the PaaS requirements which is not a trivial task. Serverless computing takes a big step in this manner. Users only have to register their functions and then receive the credentials to invoke the functions.

\subsection{Affordable scalability}
Another promise of the cloud computing is the ability for customers to deploy the functionalities of their applications without worrying about the scalability of the execution infrastructure or platform. The scalability is a direct result of the auto-scaling nature of these services i.e. per request the service invokes a copy of the requested function and there is virtually no bound for the number of concurrent requests. Each invocation is assigned to the most feasible and available resource for execution.

The affordability of serverless services is mainly due to the reduced costs of the providers. There is two main reasons for the more reduced costs, (1) resource multiplexing and (2) infrastructure heterogeneity. Resource multiplexing leads to higher utilization of available resources. For example, consider the case in which an application has one request every minute and it takes milliseconds to complete each request. In this case, the mean CPU usage is very low. If the application is deployed to a dedicated machine then this is highly inefficient. Many similar applications could all share that one machine. Infrastructure heterogeneity means that the providers also can use their older machines that are less attractive for other direct services to reduce their costs (mainly because the execution environment is a black-box for the customer).

It is important to note that the affordability of serverless services depends on the usage scenario i.e. in some cases renting a virtual machine is cheaper than using serverless's pay-as-you-go model. Lin et al. \cite{lin2020serverless} developed an economic model to establish a tradeoff between pricing of virtual machines and serverless services. Their study characterizes scenarios that the customers can benefit from switching to serverless services.

\subsection{New market places} 
With the advent of modern operating systems for mobile devices such as Android or iOS, various market places for applications, that are specifically designed for those operating systems, have emerged such as Google Play Store \cite{googleplay} and Apple's App Store \cite{appstore}. Such a scenario is already appearing for serverless paradigm i.e. with the growth in the popularity of serverless computing, new market places for functions has emerged. In these types of markets, developers can sell their developed functions to others. Every generalized or domain-specific functionality can be bought or offered in those markets. For example, a software developer may need a geospatial function that checks whether a point resides inside a geospatial polygon. They could buy such functions from those markets. AWS Serverless Application Repository \cite{serverlessrepo} is an example of such a capability. \cite{spillner2019quantitative} presents a quantitative analysis of functions available inside AWS Application Repository.

The competition forced by the economics of these markets will lead to high-quality functions i.e. both from the perspective of code efficiency, cleanness, documentation, and resource usage. The function markets may present buyers with a catalog for every function which shows the resource usage of the function and prices it incurs per request. Thus, the buyer can choose from many options for a specific task.

\begin{table*}[t]
    \centering
    \caption{Serverless application domains}
    \label{tab:the_table}
    \begin{tabular}{|l|c|c|}
        \hline
        
        \textbf{Application} & \textbf{Main reason} & \textbf{Assessment} \\\hline\hline
        Real-time Collaboration and Analytics & Auto-scaling feature & Promising\\ \hline
        Urban and Industrial Management Systems & Pricing model& Promising\\ \hline
        Scientific Computing & Lower deployment overhead & Fair\\ \hline
        Artificial Intelligence and Machine Learning & Pricing model& Fair\\ \hline
        Video Processing and Streaming  & Lower deployment overhead & Fair\\ \hline
        System and Software Security & Auto-scaling feature & Promising\\ \hline
        Internet of Things (IoT) & Auto-scaling feature & Promising\\ \hline
        E-commerce, Banking and Blockchains & Auto-scaling feature & Fair\\ \hline
    \end{tabular}
\end{table*}

\section{Applications}
\label{Sec:application}
Many real-world serverless applications have been proposed in the literature during the past few years. We categorize these applications into eight domains. Table \ref{tab:the_table} lists the application domains, their related papers, the main reason for migration stated in the papers, and assessments based on the arguments made by the authors, the obtained results, and the challenges they reported. In what follows, we survey these application domains, in detail.

\subsection{Real-time collaboration and analytics}
The stateless nature of serverless services makes them an attractive platform for real-time collaboration tools such as instant messaging and chatbots. Yan et.al., \cite{yan2016building} proposed an architecture for chatbot on OpenWhisk \cite{openwhisk}. An XMPP-based serverless approach for instant messaging is also introduced in \cite{saint2018serverless}. Real-time tracking is another example of collaboration tools that are very suitable for serverless services as these applications are not heavily dependant on the system's state. Anand et.al., \cite{anand2019real, anand2019low} proposed two real-time GPS tracking methods on low-power processors.

Serverless services are also utilized for data analytics applications \cite{nastic2017serverless}. In these applications, various sources stream real-time data to a serverless service. The service gathers, analyses, and then represents the data analytics. The auto-scaling feature of serverless computing makes the handling of concurrent massive data streams, possible.  Müller et. al \cite{muller2020lambada} proposed Lambada which is a serverless data analytics approach that is one order of magnitude faster and two orders of magnitude cheaper compared to commercial Query-as-a-Service systems.

\subsection{Urban and industrial management systems}

The pay-as-you-go model of serverless services paved the way for the introduction and implementation of various budget-restricted urban and industrial management systems.  Al-Masri et.al., \cite{al2018serverless} presented an urban smart waste management system. Hussain et.al., \cite{hussain2019serverless} proposed a serverless service for oil and gas field management system. An implementation of a serverless GIS platform for land valuation is presented in \cite{mete2021imp}. 

The distributed nature and auto-scaling feature of serverless services make it an apt choice for smart grids. Zhang et. al,  \cite{zhang2021serverless} proposed event-driven serverless services to handle SCADA/EMS failure events. A distributed data aggregation and analytics approach for smart grids is proposed in \cite{huang2021serverless}. Serverless services have been also utilized for urban disaster recovery applications. Franz et.al., \cite{franz2018reunifying} proposed a community formation method after disasters using serverless services. Another similar approach is also proposed in \cite{coleman2017speeding}.

The migration toward serverless paradigm seems a reasonable choice for this domain of applications, especially, for public sector services or for developing countries due to its lower deployment overheads and also its pay-as-you-go pricing model.


\subsection{Scientific computing}
It has been debated in \cite{hellerstein2018serverless} that serverless computing is not an attractive alternative for scientific computing applications, albeit, many studies have focused their attention toward serverless services for those applications. We believe disagreement lies in the fact that the range of scientific computing and its applications are vast and there are certainly some areas in this domain for which the utilization of serverless services is feasible. 

Spillner et.al., \cite{spillner2017faaster} argue that serverless approaches provide a more efficient platform for scientific and high-performance computing by presenting various prototypes and their respective measurements. This idea is also echoed in \cite{chard2019serverless} where high-performance Function-as-a-Service is proposed for scientific applications. A serverless tool for linear algebra problems is proposed in \cite{shankar2018numpywren} and a case for matrix multiplication is presented in \cite{werner2018serverless}. Serverless paradigm is harnessed for large-scale optimization in \cite{aytekin2019harnessing}. A serverless case study for scientific workflows is discussed in \cite{malawski2016towards}. 

Serverless approaches have been also used in DNA and RNA computing \cite{lee2019dnavisualization, hung2019rapid}. Niu et. al, \cite{niu2019leveraging} utilized the potentials of serverless paradigm in all-against-all pairwise comparison among all unique human proteins. On-demand high-performance serverless infrastructures and approaches for biomedical computing are proposed in \cite{kumanov2018serverless}.

Scientific applications that require extensive fine-grained communication are difficult to support with a serverless approach, whereas those that have limited or coarse-grained communication are good candidates. Also, note that scientific computations with time-varying resource demands will benefit from migrating to a serverless paradigm.

\subsection{Artificial intelligence and machine learning }
Machine learning in general and neural network-based learning, in particular, are currently one of the most attractive research trends. The suitability of the serverless paradigm for this domain has received mixed reactions both from research and industrial communities. For example, it has been argued that deep learning functions are tightly coupled (they require extensive communication between functions), and also these functions are usually compute and memory intensive, as such, the paradigm is not promising for these applications \cite{feng2018exploring}. Nevertheless, it has been discussed that deep neural networks can benefit from serverless paradigms as they allow users to decompose complex model training into several functions without managing virtual machines or servers \cite{xu2021lambda}. As such, various such approaches have been proposed in the literature. A case of serverless machine learning is discussed in \cite{carreira2018case}. Ishakian et al., \cite{ishakian2018serving} discussed various deep learning models for serverless platforms. Neural network training of serverless services is explored in \cite{feng2018exploring}. Also, a pay-per-request deployment of neural network models using serverless services is discussed in \cite{tu2018pay}. A prototype serverless implementation for the estimation of double machine learning models is presented in \cite{kurz2021distributed}. A distributed machine learning using serverless architecture also discussed in \cite{wang2019distributed}.



Christidis et al., \cite{christidis2020enabling} introduce a set of optimization techniques for transforming a generic artificial intelligence codebase to serverless environments. Using realistic workloads of the UK rail network, they showed that by wielding their techniques the response time remained constant, even as the database scales up to 250 million entries. A serverless framework for the life-cycle management of machine learning-based data analytics tasks is also introduced in \cite{bhattacharjee2019stratum}.

The viability of serverless as a mainstream model serving platform for data science applications is studied in \cite{wu2021serverless}. The authors presented several practical recommendations for data scientists on how to use the serverless paradigm more efficiently on various existing serverless platforms. 

\subsection{Video processing and streaming}
Serverless approaches have been proposed for video processing. Sprocket \cite{ao2018sprocket} is a serverless video processing framework that exploits intra-video parallelism
to achieve low latency and low cost. The authors claim that a video with 1,000-way concurrency using Amazon Lambda on a full-length HD movie costs about \$3 per hour of processed video. In another interesting work, a serverless framework
for auto-tuning video pipelines discussed in \cite{romero2021llama}. It achieves 7.9 times lower latency and 17.2 times cost reduction on average compared to that of serverful alternatives. In  \cite{risco2021gpu} GPU processing power is harnessed in a serverless setting for video processing. Zhang et al., \cite{zhang2019video} present a measurement study to extract contributing factors such as the execution duration and monetary cost of serverless video processing approaches. They reported that the performance of video processing applications could be affected by the underlying infrastructure.

Serverless video processing and broadcasting applications have gained much traction both from industrial and research communities during the COVID-19 pandemic. A live media streaming in a serverless setting is presented in \cite{konstantoudakis2021serverless}. A serverless face-mask detection approach is discussed in \cite{wang2021wearmask}. Serverless paradigm also has been utilized in video surveillance applications. Elordi et al. \cite{elordi2021demand} proposed an on-demand serverless video surveillance using deep neural networks.

\subsection{System and software security}
The power of serverless computing has been leveraged for providing security for various software systems and infrastructures. A mechanism for securing Linux containers has been proposed in \cite{bila2017leveraging}. Serverless services have also been utilized for intrusion detection. StreamAlert \cite{naglieri2017streamalert} is a serverless, real-time intrusion detection engine built upon Amazon Lambda. Birman et al. \cite{birman2020cost} presented a serverless malware detection approach using deep learning.

Serverless approaches have been also used for ensuring data security. A method for automatically securing sensitive data in the public cloud using serverless architectures has been introduced in \cite{fonseka2019system}. Hong et al. \cite{hong2018go} presented six different serverless design patterns to build security services in the cloud. 

We believe that the serverless approach has great potentials to improve the security of systems and services. This is due to various reasons:
\begin{enumerate}
    \item Security threats are often ad-hoc in nature. In these cases, the pay-as-you-go pricing leads to reduced costs.
    
    \item Some of the attacks exhibit sudden traffic bursts. The auto-scaling feature of the serverless services facilitates the handling of such a scenario. 
    
    \item Attackers may conduct widespread attacks interrupting various components and infrastructures of the victim. Serverless functions are standalone in the sense that the functions can be executed in various execution environments.
\end{enumerate}


In the opposite direction, serverless approaches have been used to develop a botnet \cite{wu2018slbot}. We think that preventing attackers from using serverless infrastructures to conduct these types of attacks is an important issue that needs to be addressed.

\subsection{Internet of Things (IoT)}
The serverless computing paradigm has been exploited for various IoT domains. Benedetti et al. \cite{benedetti2021experimental} conducted various experiments on IoT services to analyze the aptness of different serverless settings. Using a real-world dataset, authors of \cite{wang2020supporting} showed that a serverless approach to manage IoT traffic is feasible and utilizes fewer resources than a typical serverful approach. Cheng et al. \cite{cheng2019fog} propose a serverless fog computing approach to support data-centric IoT services. A smart Internet of Things (IoT) approach using the serverless and microservice architecture is proposed in \cite{herrera2018smart}. A serverless body area network for e-health IoT applications is presented in \cite{regan2019serverless}. A serverless IoT platform for smart farming is introduced in \cite{trilles2020iot}. Serverless paradigms also have been utilized for coordination control platforms for UAV swarms \cite{hu2020hivemind}.

In another research direction, Presson et al. \cite{persson2017kappa} introduced a flexible and intuitive serverless platform for IoT. A decentralized framework for serverless edge computing in the Internet of Things is presented in \cite{cicconetti2020decentralized}. The objective of the paper is to form a decentralized FaaS-like execution environment (using in-network executors) and to efficiently dispatch tasks to minimize the response times. In another interesting work, George et al. introduced Nanolambda \cite{george2020nanolambda} which is a framework that brings FaaS to microcontroller-based IoT devices using a Python runtime system. Amazon's Greengrass \cite{greengrass} also provides a serverless edge runtime for IoT applications.

It is reasonable to confer that serverless services can act as feasible back-ends for IoT applications that have infrequent and sporadic requests. For the scenarios where rapid unpredictable surges of requests emerge, serverless services can conveniently handle requests as they can auto-scale rapidly.

\subsection{E-commerce, banking and crypto-currencies}
The inherent scalability of serverless services has enticed few e-commerce and banking applications. A serverless implementation of a core banking system is presented in \cite{kohler2021serverless}. Goli et al. \cite{goli2020migrating} present a case study of migrating to serverless in the FinTech industry. Huy et al. \cite{huy2021crypto} implemented a crypto-currency tracking system based on OpenWhisk. 


\begin{figure*}[t]
\centering
\includegraphics[width=6in]{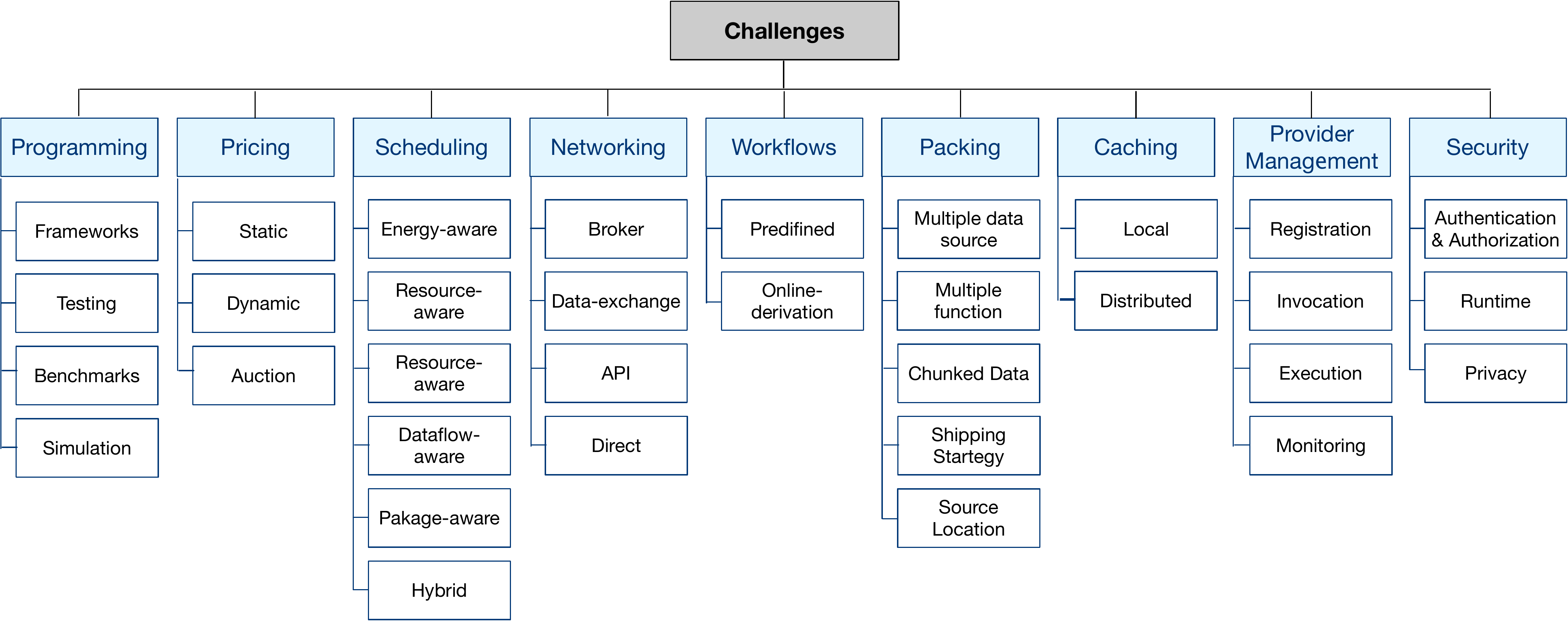}
\caption{An overview of the challenges discussed in this paper}
\label{fig:chart}
\end{figure*}

\section{Challenges}
\label{sec:challenges}

In this section, we summarize and discuss the challenges faced by the application domains surveyed in Sec \ref{Sec:application}. We categorize those challenges into nine topics and survey the existing solutions for each of them. Figure \ref{fig:chart} depicts an overview of these topics. We also present the areas that need further attention from the research community and identify open problems.

\subsection{Programming, modeling, testing, and debugging}
As the topic of serverless computing is relatively new, its development tools, concepts, and models are not rich enough. This poses a great challenge for software developers. The lack of proper modeling paradigms leads to non-unified development approaches which will reduce the quality of the code and also complicate collaborations of developers in long term. To remedy this shortcoming, Perez et al. \cite{perez2019programming} propose a programming model and middleware for serverless computing applications. The focus of the paper is limited to file processing applications. A formal model for serverless computing using lambda calculus is presented in \cite{gabbrielli2019no}. Also, model-based analysis of the serverless application is discussed in \cite{winzinger2019model}.

To enhance the adoption of a novel paradigm either new programming frameworks should be developed or existing programs on legacy frameworks should be ported to the new paradigm. Zhang et al. \cite{zhang2020kappa} introduced Kappa which is a programming framework for serverless computing. It facilitates and simplifies parallel programming on serverless platforms i.e. the programmer can write ordinary Python code and Kappa transforms the code and executes it on a serverless platform using parallel lambda functions. Pywren \cite{jonas2017occupy} enables programmers to directly run their existing Python codes on Amazon's Lambda platform.

Debugging and testing tools are integral parts of any software development approach. Serverless computing is not an exception. Few papers have elaborated on this. Lenarduzzi et al. \cite{lenarduzzi2020serverless} reflect expert views on the issues and challenges that need to be investigated.  \cite{manner2019troubleshooting} proposes a combined monitoring and debugging approach for serverless functions. Integration testing of serverless functions is discussed in \cite{winzinger2020applicability}.

Another important tool to test the applicability and performance of any new idea is benchmark suites. Several benchmark tools have been developed for serverless applications and service providers during the past few years. A rich set of benchmark suites for serverless infrastructures is presented in \cite{gan:asplos:2019:microservices}. Yu et al. \cite{yu2020characterizing} introduced \textit{ServerlessBench}, an open-source benchmark suite that includes many test cases to explore important metrics such as communication efficiency, startup latency, and overall performance. FaaSdom \cite{maissen2020faasdom} is another benchmark suite for serverless computing platforms that also integrates a model to estimate budget costs for deployments across the supported providers. Another such benchmarking tool is PanOpticon \cite{somu2020panopticon} which also provides an array of configurable parameters and gives out performance measurements for each selected configuration and platform.

Simulation tools are also important for rapid modeling of real-world situations or testing new ideas and concepts. Mahmoudi et al. \cite{mahmoudi2021simfaas} introduced SimFaaS which is a performance simulator for serverless platforms. It simplifies the prediction of various serverless computing key performance metrics.

\subsection{Pricing and cost prediction}

Many big technology companies now offer serverless computing services with different specifications and prices. As the popularity of serverless services increases, the number of companies and their options for pricing will grow. Many factors affect the price offered by each company. These factors range from the company's revenue strategy to the platform it uses and the energy prices (based on the region or the time of day during which the function execution occurs). For example, a request that gets to a server at 2 a.m. in winter typically costs lower compared to that of the same request with the same resource consumption at 2 p.m. workday in the summer. Another factor is the load level that is imposed on the provider at that moment i.e. whether the provider nearing its peak power demand or not. Peak demand prices are reported to be 200-400 times that of the nominal rate \cite{nghiem2011green}. Consequently, the contribution of peak charge in the electricity bill for a service provider can be considerable, e.g., from 20\% to 80\% for several Google data centers \cite{xu2014reducing}. The price offered by the competitors is also a key decision factor. Various pricing models have been proposed for cloud computing in general \cite{al2013cloud} that are not directly applicable to serverless service. Extracting a pricing model for service providers is a challenging issue that should further be studied by the research community.

The pricing problem is also important for customers. As discussed above, the diversity of the prices will lead to a competitive environment between service providers. Thus, the customer can choose between various price options in an online manner to reduce the costs. In this way, customers put their functions on multiple serverless services (or ship it instantly) and then based on the online available prices, the software decides to place the request to the most affordable service provider. The ultimate goal of customers is to reduce their costs while maintaining the quality of service. Note that, one of the important factors in determining the quality of service is the response time.

Finding an optimal or a sub-optimal pricing strategy with multiple providers and customer constraints is a challenging issue that must be addressed in research studies. A similar notion has been extensively discussed for cloud computing in general, where supply and demand are considered in extracting dynamic pricing models \cite{xu2013dynamic}.

It is noteworthy to mention that, the nature of serverless services makes online pricing more feasible compared to that of other cloud services. In those services such as IaaS, the cost of moving and maintaining several virtual machines in various service providers is higher compared to that of function placement in several serverless providers. This enables the scenario in which function placement can be done using auctions as discussed in \cite{bermbach2020towards}.

Currently, major serverless providers only offer static pricing. This makes the task of predicting the customer costs straightforward as the costs only depend on the resource usage rather than other somewhat harder to predict parameters (such as time of use, etc.). Thus, several research studies have focused on either predicting or modeling resource usage of serverless functions. Using Monte Carlo simulation, Eismann et al. \cite{eismann2020predicting} proposed a methodology for the cost prediction of serverless workflows. They showed that the proposed approach can predict the response time and output parameters of a function based on its input parameters with an accuracy up to 96.1\%. Cordingly et al. \cite{cordingly2020predicting} introduced a tool for accurate performance predictions with error percentage below 4\%. An analytical model using a heuristic algorithm is presented in \cite{lin2020modeling}. The results show accuracy up to 98\%. The latter approach is easier to implement and faster to attain as it only executes a greedy algorithm. Overall, it seems that the problem of predicting resource usage of serverless functions is tractable with high accuracy.


\subsection{Scheduling}
\label{sec:scheduling}


Invocation requests are sent to the provider either by customers' applications or other functions. These requests often have predefined deadlines. This is especially of great importance for real-time, latency-sensitive, or safety-critical applications. The provider must schedule where (i.e. which computation node) and when to execute the functions such that it conforms with the deadlines while considering other system-related criteria such as energy consumption, or resource utilization. There are various strategies that the scheduler could adopt. In what follows, we summarize those strategies. Note that, the real-world schedulers may adopt one or more of these strategies.
\\

\noindent \textbf{Energy-aware scheduling:}
The main idea in this type of scheduling is to put inactive containers or the execution environment in a hibernate mode (or cold-state mode) to reduce energy consumption. The transition from cold-state to active mode incurs delays in the execution of invoked functions which may go beyond the deadlines defined by the customer\footnote{By our estimates on a laboratory installation of OpenWhisk, the cold start latency can range from 2 to 6 seconds; by contrast, real-world open-source function executions usually take millisecond scale}. Thus, in these approaches, the executions are scheduled so that the number of such transitions minimizes. For example, Suresh et al. \cite{suresh2020ensure} introduced ENSURE which is a scheduler for serverless applications. To prevent cold starts, it proactively reserves a few additional containers in a warm state which can smoothly handle workload variations. Using a theoretical model they tried to minimize the amount of additional capacity while ensuring an upper bound for the request latency. Fifer \cite{gunasekaran2020fifer} also uses a similar approach i.e. it proactively spawns containers to avoid cold-starts.

An energy-aware scheduler also can take advantage of delaying non-latency-sensitive tasks (such as background or maintenance tasks) to reduce overall energy consumption. We think introducing execution scheduling for a mixture of latency-sensitive and non-latency-sensitive functions can be a good direction for future research studies.
\\

\noindent \textbf{Resource-aware scheduling:}
Serverless applications are diverse, so are their resource consumption patterns. For example, a scientific computing function is usually CPU-intensive while an analytic application is often memory-intensive. Co-locating many CPU-intensive functions in a physical node leads to resource contention and may incur delays to the execution of those functions. This is also true for other types of computing resources such as memory, disk, and network. Resource-aware schedulers place functions to the computation nodes so that they can provide resource requirements of the functions in a timely manner. FnSched \cite{suresh2019fnsched} falls into this category of serverless schedulers. it first categorizes functions based on their CPU usage. It then reduces CPU contention between co-located functions by dynamically regulating their CPU shares at runtime. In another approach, Kim et. al. \cite{kim2020automated} proposed a dynamic CPU capping method along with a group-aware scheduling algorithm for serverless computing which can effectively reduce CPU contention between functions inside computation nodes. HoseinyFarahabady et. al. \cite{hoseinyfarahabady2017model} proposed a prediction tool to estimate the rate of event streams inside providers with the end goal of reducing CPU contention and improving QoS. A QoS-aware resource allocation scheme that dynamically scales by predicting the future rate of incoming events for serverless environments is also introduced in \cite{hoseinyfarahabady2017qos}. In another direction, Fifer \cite{gunasekaran2020fifer} conducts offline profiling to calculate the expected execution time of functions and balances the load adaptively. 
A framework that uses Bayesian Optimization to find the optimal configuration for serverless functions is presented in \cite{akhtar2020cose}. It uses statistical learning techniques to gather samples and predict the cost and execution time of a serverless function across unseen configuration values. The framework uses the predicted cost and execution time, to select the best configuration parameters for running a single or a chain of functions while satisfying customer objectives. 
\\

\noindent \textbf{Workflow-aware scheduling:} Serverless applications usually require the execution of several functions to handle their tasks. These stateless small discrete functions are chained together and orchestrated as serverless workflows. We describe this in Sec. \ref{sec:workflow} comprehensively. By knowing the chain of function invocations, schedulers can speculate the next functions in the chain to forecast, schedule, and provision in advance e.g., prepare a warm container for the execution. Xanadu \cite{daw2020xanadu} uses such speculation to reduce the overheads and delays up to 10 times compared with OpenWhisk. Sequoia \cite{tariq2020sequoia} and Archipelago \cite{singhvi2019archipelago} also adopt variations of this strategy. \cite{singhvi2019archipelago} uses DAG-structure\footnote{Directed Cyclic Graphs (DAG) is usually used to show dependency between tasks in scheduling algorithms} to predict the size of worker pools and thus achieve low scheduling overheads for request execution. Sequoia allows prioritizing, scheduling, and queuing of function chains, or functions within workflow chains in order to improve Quality-of-Service (QoS). 

We think that this area has the potential for further studies. For example, probabilistic serverless DAGs discussed in \cite{lin2020modeling} can be leveraged to further improve workflow-aware scheduling methods.
\\

\noindent \textbf{Dataflow-aware scheduling:}
Although in an ideal serverless setting the functions are stateless and do not depend on any external data sources, in practice, this is usually not the case. For example, in many machine learning applications, the dependency on external sources is high.  The scheduler thus should take the availability of the data or the data serving delays into account. Hunhoff et al. \cite{hunhoff2020proactive} introduced \textit{freshen} which allows developers or providers to proactively fetch data along with other runtime reuse features to reduce overheads when executing serverless functions. The scheduling policy of Cloudburst \cite{sreekanti2020cloudburst} also prioritizes data locality. Rausch et al. \cite{rausch2021optimized} presented a domain-specific dataflow-aware serverless scheduler for edge computing.  

We further investigate the data dependency in Sec. \ref{sec:packing}. Data caching is also of great importance to reduce the delay imposed by data dependency, we discuss this in Sec. \ref{sec:caching}.
\\

\noindent \textbf{Package-aware scheduling:}
In some of the current serverless technologies e.g., OpenLambda, the computation node should fetch and install application libraries and dependent packages declared by function upon receiving an invocation request. This obviously takes some time and delays the execution. Amumala et al. \cite{aumala2019beyond} proposed a package-aware scheduling scheme that addresses this issue. Other serverless platforms are usually designed such that the customers themselves incorporate those package during the registration phase. 

This area of research also has potentials for further investigations. For example, machine learning approaches could be utilized to predict future needed libraries based on currently installed libraries similar to the next basket recommendation problem \cite{bai2018attribute}.
\\

\noindent \textbf{Hybrid scheduling:}
Hybrid virtual-machine/serverless scheduling also has gained attention recently. The idea is to have the best of all worlds scenario; a central scheduler decides to place requests to a private virtual machine or to a serverless provider. Spock \cite{gunasekaran2019spock} is a hybrid scheduling mechanism that flattens request peaks using VM-based auto-scaling. Skedulix \cite{das2020skedulix} is also a hybrid scheduler with the objective of minimizing the cost of using public cloud infrastructures while conforming with user-specified deadlines.

\subsection{Networking, sharing and intra-communications}
A serverless software is typically a composition of many functions that work together to provide the desired functionality. To attain this, the functions need to somehow communicate with each other and share their data or their state. In other cloud services, this is attained through network addressing. For example in IaaS, each virtual machine can send and receive data through point-to-point networking using network addresses. 

Functions intra-communication and network-level function addressing in serverless platforms are challenging. Functions in serverless services have characteristics that must be considered to be able to introduce some kind of addressing scheme for them:

\begin{itemize}
\item Due to the auto-scaling nature of serverless computing, at any given time there may be several running invocations of the same function inside various computation nodes around the world. This rules out the possibility of addressing based on function name or location.

\item The functions are often short-lived. The short life span of the functions means that any addressing scheme should be fast enough to track the rapid changes of the system's entire state.

\item With the growth in the usage of serverless services, the number of copies of functions that are being deployed will grow drastically. Thus, the proposed addressing space should be scalable enough to be able to handle that volume of functions. 

\end{itemize}

Even with a proper addressing scheme, intra-communication between functions is still challenging. There are several possible approaches:

\begin{enumerate}
\item Intermediate functions or external coordinators that serve as brokers between functions that are suggested in \cite{fouladi2019laptop}. The same idea is also utilized in  \cite{wang2020infinicache} where proxies act as coordinators. However, it has been argued in \cite{jonas2019cloud} that this burdens extreme overhead on the infrastructure.  Performance evaluations of this approach that reveal the bottlenecks and further investigation and optimization of contributing factors, is a good direction for related research studies.

\item Communication through data exchange is another approach. This is achievable either through data exchange mechanisms that rely on cloud storage to pass the data such as that of \cite{muller2020lambada} and \cite{perron2020starling}, or using distributed message queues such as Amazon's Kinesis Data Streams \cite{kinesis}. There is also the possibility of tailored VM-based resources exchange mechanisms \cite{pu2019shuffling, wu2020autoscaling}. This type of communication imposes an order of magnitude higher latency than point-to-point communications \cite{wawrzoniak2021boxer}.  Introducing new fast protocols and methods for data exchange that is specifically designed for function communication in serverless environments needs researchers' attention. 

\item Communicate through APIs using stateless communication schemes (or protocols) such as REST or SOAP. These protocols are vastly used over the Internet and seem to be good candidates. Such an approach is introduced by \cite{snsdk} using specialized APIs and network protocols.

\item Direct network sockets over TCP/IP is also another approach that is introduced in \cite{wawrzoniak2021boxer} and \cite{thomas2020particle}. The idea is to enable generalized networking capabilities for functions rather than facilitating only function-to-function communications. Wawrzoniak et al. \cite{wawrzoniak2021boxer} proposed such a method over TCP/IP. Their benchmark shows a sustained throughput of 621 Mbit/s and a round-trip latency of less than 1 ms. In another related approach, Thomas et al. introduced Particle \cite{thomas2020particle} which is an optimized network stack for serverless settings. It improves application runtime by up to 3 times over existing approaches.

\end{enumerate}




\subsection{Serverless workflows}

\label{sec:workflow}



As discussed earlier in this paper, a serverless application is a composition of various functions working in coordination with each other to accomplish the desired tasks. Rarely, we have applications that are composed of a single function, instead, usually, there are many interdependent functions, processing and passing data to each other and send back the result to the application. For example, real-world implementation of an online social network (with functionalities similar to Facebook \cite{facebook}) on Amazon's Lambda infrastructure has around 170 functions \cite{adzic2017serverless}. These small functions are orchestrated into high-level workflows to form serverless applications.

In each application, functions are executed in various sequences. For example, users \textit{sign up} in the application, view latest \textit{products}, click on the \textit{add-to-cart} button and \textit{check out}. These 4 functions are executed in a sequence. Obviously, this is not the only execution sequence in the application since the user may have already signed up and just needs to \textit{sign in}. There are many other possibilities for the sequences of functions. As we discuss in this paper, knowledge of these sequences (or chains) plays a key role in improving the performance of serverless services. Providers can use this knowledge to pre-fetch, prepare and optimize functions to reduce costs and serve customers with better performance.

\begin{figure}[t]
\centering
\includegraphics[width=2.65in]{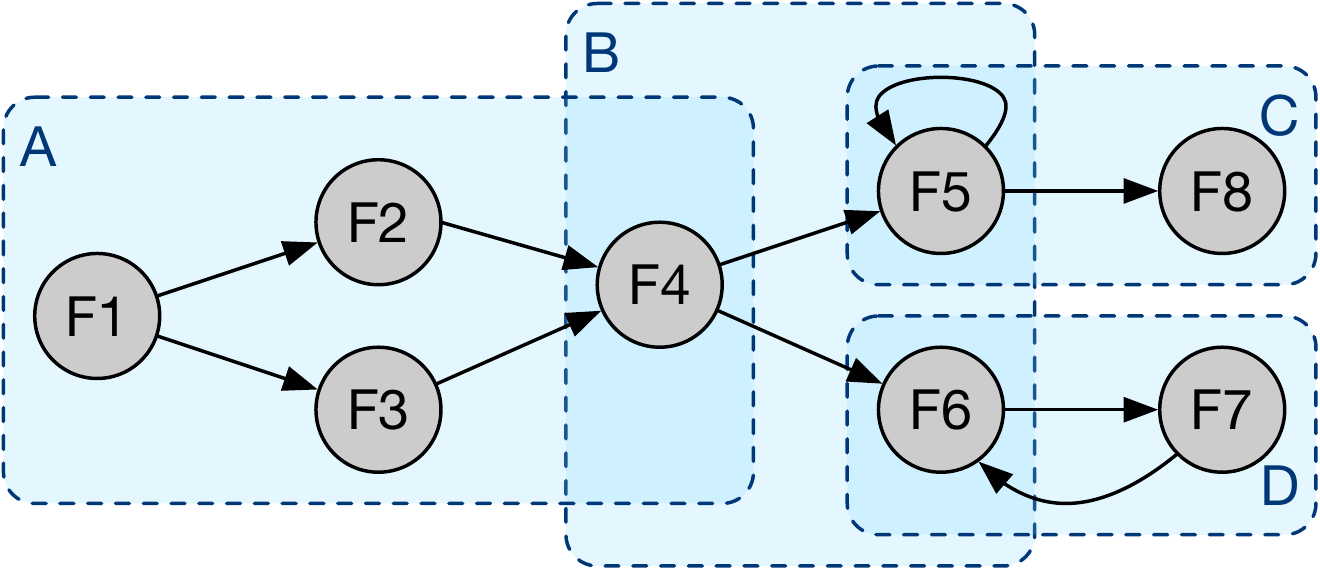}
\caption{An example of a task graph for a serverless application. This application has 8 functions where the start function is F1. Four different types of structures in task graphs are shown i.e. (A) a parallel execution, (B) a branch, (C) a self-loop, and (D) a cycle.}
\label{fig:depgraph}
\end{figure}

From each serverless application, a directed task graph can be derived where the nodes are functions and edges show the dependency or precedence of the execution of one function to another. Figure \ref{fig:depgraph} shows an example of such a graph. The application depicted in the figure has 8 functions (F1 to F8). A directed edge from F1 to F2 means that the execution of F2 depends on the execution of F1. For example, users must \textit{sign up} to be able to \textit{check out}. The figure shows different types of sub-structures in the task graph i.e. parallel execution, cycle, self-loop, and branch. Each of these structures must be considered and investigated inside providers to improve the system's overall performance as studied in \cite{lin2020modeling}.


There are two main approaches to attain workflows for a serverless application:

\subsubsection{Predefined workflows} 
In this approach, the application developer submits a file that contains the workflow or defines it through an interface. The provider receives and processes the workflow for further actions. Many of the existing serverless providers offer this service. AWS Step Functions \cite{awsstep}, Azure Durable Functions \cite{azuredurable}, Alibaba Serverless Workflow \cite{alibabaworkflow}, and Google Cloud Composer \cite{gcomposer} are examples of such a service.

Various approaches exist to define and specify workflows for serverless platforms. Amazon uses States Language \cite{asl} which is a language that enables users to define state machine and determine states transitions. Microsoft Azure uses standard programming language of choice (such as JavaScript, Python, C\#, or PowerShell) code to represent workflows \cite{burckhardt2021serverless}. Google's composer service enables users to use Python to define workflows. To enable a platform-independent language for various existing platforms, Ristov et al. \cite{ristov2021afcl} introduced AFCL which is a  generic language for serverless workflow specification that can be translated to multiple platforms e.g., AWS Lambda or IBM Cloud Functions. Triggerflow \cite{lopez2020triggerflow} is another interesting approach to compose event-based workflows for serverless systems.

Wen et al. \cite{wen2021empirical} conducted an empirical study of these serverless workflow services. They thoroughly investigated the characteristics and performance of these services based on real-world workloads. In another related study, the performance of sequential workflows under various platforms is discussed in \cite{burckhardt2021serverless}. Domain-specific serverless workflows also have been proposed in the literature to accelerate serverless application development in that domain. SWEEP \cite{john2019sweep} which is a utility to define, execute and evaluate workflows of scientific domains, falls into this category.


\subsubsection{Detection and extraction during execution}  
In this approach, a workflow graph is formed inside the service provider based on the activities of the application. This approach is simpler for the developer since it does not need any further workflow definition tools or services. It is also more flexible in the sense that it extracts the chains of execution based on the actual behavior of the application (these chains are described as \textit{implicit} chains in \cite{daw2020xanadu}). In large applications with hundreds or even thousands of functions, there are many paths and sometimes it is not trivial to predict the exact behavior of the users to define them. Exploiting them during execution leads to more accurate estimations. This approach is also more robust than the other since the workflows defined by the users could be faulty or less accurate. We think this topic is a good direction for future research studies.

\subsection{Packing}
\label{sec:packing}
The main incentive for the migration toward a serverless service is its ability to auto-scale itself by executing copies of customers' functions and assign each request to those copies. Schedulers select the physical node to execute the functions based on various strategies that we discussed in Sec \ref{sec:scheduling}. In many real-world scenarios, functions have certain dependencies on a single remote data source or even several data sources. The data must be shipped from those sources to the computation nodes. This leads to orders of magnitude slower function execution \cite{jonas2019cloud} mainly because of higher traffic inside the provider which increases the latency and reduces the overall performance of the system. Thus, it seems wise to ship the function as near as possible to the data i.e. ``pack'' functions with data. 

To attain a robust approach for packing of functions with data, Zhang et al. \cite{zhang2019narrowing} proposed Shredder which is a multi-tenant cloud store that allows serverless functions to be performed directly within storage nodes (i.e. packing functions to storage). However, this approach leads to complexity in provisioning and utilization as the storage and compute resources are co-located. To overcome this issue \cite{bhardwaj2020adaptive} propose an approach that enables users to write storage functions that are logically decoupled from storage. The storage servers then choose where to execute these functions physically using a cost model. 

It is also possible to pack data and functions together. Shashidhara \cite{shashidharalambda} proposed Lambda-KV, which is a framework that aggregates and transforms compute and storage resources into a single entity to enable data locality (i.e. packing functions and storage together). This approach is not suitable for the scenarios in which the data changes with high frequency e.g., IoT sensor data readings in certain scenarios. However, the method is applicable to many transaction-based applications such as that of e-commerce and banking applications or that of background applications such as video processing tasks.

Choosing between the two approaches i.e. packing data to functions or the other way is highly dependant on the application domain. As such, Kayak \cite{youship} is proposed that adaptively chooses between shipping data to functions (i.e. packing storage to functions) or vice versa. It maximizes throughput while meeting application latency requirements. There are further considerations for packing of functions that must be taken into account:
\\

\noindent \textbf{Multiple data sources and a single function:}  There are some scenarios in which a function consumes various data sources. The basic idea would be to ship the data sources together and then pack the function with those data sources. However, this may not be feasible due to various reasons: (1) one or more of the data sources are already near other functions that consume the data, (2) the movement is not physically possible due to the lack of sufficient storage, (3) the movement is not feasible since the number of times that functions access the data is considerably low. Finding an optimal position for the function based on the distance between the physical location of the function and its data sources on the network topology is an interesting problem that needs to be addressed.
\\

\noindent \textbf{Multiple functions and a single data source:} This case is actually simpler. In this case, multiple functions are packed together with the data source in one machine. Fasslets \cite{shillaker2020faasm} is well-suited for this scenario. It provides isolation abstraction for functions based on software-fault isolation (SFI) while enabling memory regions to be shared between functions in the same machine. Photon \cite{dukic2020photons} tries to co-locate multiple instances of the same function within the same runtime to benefit from the application state and data. 
\\

\noindent \textbf{Chunked data:} The packing of function and data can be done with chunks of data instead of the whole data. For example, for a function that queries customers' tables of a company's database, that specific table is important and can be packed with the function instead of the whole database.
\\

\noindent \textbf{Evolutionary vs revolutionary movement of data:}
As mentioned above, there are scenarios in which data must be moved toward the function. This can be done in evolutionary or revolutionary modes. In the former mode, the chunks of data are moved based on requests from the function, the movement is done incrementally. This may lead to inconsistency in the data which must be taken care of by the provider. In the latter, the data is moved altogether.
\\

\noindent \textbf{Source location:} The relative geographical location of the request to the function also may play a role. Packing data and functions together and then shipping them to the nearest possible location to the requester would reduce the delays that the service faces due to network traffic. 

The packing can be done before, during, or after the first execution of a function. In the case in which the packing is done before the execution of the function, a careful manifestation of data dependency is needed to find the optimum placement of the function. To this end, Tang et al. \cite{tang2020lambdata} proposed LambData that enables developers to declare function’s data intents. In the evolutionary model, the packing is done during the execution. It can also be done after the first execution. In this case, the execution logs are inspected after the first execution and by using optimization techniques and machine learning approaches the optimal packing strategy is extracted to improve the performance of the service and to minimize costs. Note that, packing may undermine the benefits of statistical multiplexing, leading to queuing delays and inefficient resource utilization which is a good direction for future research studies.

\subsection{Data caching}
\label{sec:caching}
To avoid the bottlenecks and latencies of persistent storage, software systems utilize multiple levels of caches. This is a common practice among cloud-based applications \cite{arteaga2016cloudcache}. Utilizing caches in serverless environments is a challenging issue since functions are executed independently of the infrastructure. For example, consider a serverless customer management application. When a function requests the data of a user from the database, the platform usually caches the data to handle any further requests and to reduce the number of costly database accesses. This works perfectly in serverful services. However, in a serverless service, the next execution of the function may be assigned to another available computation node, which renders the caching useless. This is also true when multiple functions consecutively work on a chunk of data i.e. if the computation node changes, the cached data becomes expired. Without caching, the costs, overheads, and latency grow dramatically which makes serverless services infeasible. Thus, this is one of the important challenges toward the successful implementation of any serverless service. Few caching mechanisms have been proposed for serverless systems. Amazon offers ElastiCache \cite{elasticache} which is an in-memory cache and data store. It is at least 700x more expensive than Amazon's storage service (called S3). Pu et al. \cite{pu2019shuffling} proposed a method to attain a cost-efficient combination of slow storage and costly in-memory caches. InfiniCache \cite{wang2020infinicache} is another in-memory object caching system based on stateless cloud functions. CloudBurst \cite{sreekanti2020cloudburst} also proposes a caching mechanism in its architecture.  

We think this subject has the potentials for many further research studies. In designing and implementing caches, the following must be considered:
\\

\noindent \textbf{Effect of packing:} In one of the packing schemes i.e. packing function with data, the functions are shipped as near as possible to the data. This may lead to a scenario in which multiple invocations of the same function are executed in a computation node near the data. This actually reduces the complexity of caching. Instead of focusing on a system-wide cache solution, one can focus on efficient local caching mechanisms. The action of packing also tends to ship other functions that consume the data toward the vicinity of the data, and thus with proper local caching, the chance of cache hits is improved.
\\

\noindent \textbf{Effect of workflows:} The sequence upon which a batch of functions is executed also has a great impact on designing caches. In fact, in a sequential execution, the likelihood of data dependency between two or more consecutive functions is high. Thus, caching will be effective if the execution chains are considered in the local caching strategy.
\\

\noindent \textbf{Local caching vs distributed caching}
In some of the real-world scenarios of serverless computing, an efficient local caching can be feasible\footnote{Here, by local we mean a caching mechanism shared between multiple servers in a rack or possibly a cluster of racks near each other.}. However, there are cases in which the functions cannot be shipped to the vicinity of the data. In these cases, distributed caching can be utilized. 

Distributed in-memory caches often utilize distributed hash functions (DHT) to extract the location of cached data \cite{vaquero2011dynamically}.  Then, the data is routed to the requester. If the data does not reside in the distributed cache, the requester extracts the data and caches it. This works well when the cost of extracting data from its source is higher than that of getting it from the remote cache. CloudBurst \cite{sreekanti2020cloudburst} proposes a variation of this idea. It utilizes both DHT-like distributed storage along with local caching to improve the performance of serverless applications. 

Note that, using distributed caching may incur more costs. We are facing a scenario in which the function cannot be shipped to the vicinity of the data. In this scenario, caching the data in the server that executes the function may accelerate the future invocations of the same function. However, as most other functions are shipped near the data, the cost of routing the cached data to the functions compared to that of extracting it from the source directly may actually be higher. This must be considered in any distributed cache design for serverless services.
\\

\subsection{Provider Management}
The management operation inside the serverless providers is a complex and resource-demanding task. It involves many monitoring and provisioning operations for each of the infrastructures inside the provider. The controller should handle and keep track of functions' registration, invocation, and execution. Below, we discuss each operation in detail:
\\

\noindent \textbf{Registration:} Every user should be able to upload their function, select its required resources. The provider then sends back credentials for invocation. Other tracking and business aspects are handled by the provider during this step.
\\

\noindent \textbf{Invocation:} A provider receives invocation requests from applications or other functions, checks the requester's credentials, and then finds a feasible computation node and assigns the function to the node for execution. The tasks of placement, scheduling, packing, and caching are part of the responsibility of the controller which is done in collaboration with the computation nodes.
\\

\noindent \textbf{Initialization and Execution:}
To execute functions, providers often use sandbox execution environments to provide strong isolation between function instances. OpenWhisk uses containers for the execution \cite{openwhisk}, however, containers usually have isolation problems. As such, many recent studies have focused on providing serverless sandboxes with reliable isolation. Catalyzer \cite{du2020catalyzer} is an example of such efforts. It provides strong isolation with sub-millisecond startup time. FireCracker \cite{agache2020firecracker} introduces Virtual Machine Monitor (VMM) device model and API for managing and configuring MicroVM. It provides strong isolation with minimal overhead (less than 5MB of memory) and millisecond scale boot time. Unikernels in which the function is linked with a bare minimum library operating system is another approach to attain a sandbox execution environment for serverless computing \cite{tan2020towards}.
\\

\noindent \textbf{Monitoring:} Although the execution takes place inside the computation nodes, the controller should closely monitor the execution of functions to detect errors and malfunctions. It gathers the execution logs to analyze the footprints and thus improve future invocations. Various commercial monitoring approaches exist for serverless and cloud systems such as Epsagon\footnote{\texttt{https://epsagon.com/}}, Datadog\footnote{\texttt{https://www.datadoghq.com/}}, and Dynatrace\footnote{\texttt{https://www.dynatrace.com/}}. These solutions usually are restricted to basic metrics such as CPU utilization. Eismann et al. \cite{eismann2020sizeless} proposed a  resource consumption monitoring module specifically tailored for serverless platforms.

There are two approaches to attain a controller system for serverless providers; centralized or distributed. While the centralized approach is more trivial and more efficient, it may experience extreme loads and it could become the single point of failure. Distributed monitoring, on the other hand, is complex and hard to implement.

For the controller to be able to handle its responsibilities, manage resources and optimize the services, it should have an online view of the entire system. Various pieces of information contribute to the formation of this view, such as:

\begin{enumerate}
\item Information about the functions: their data dependency, the workflow, their owner, the origin of the requests, rate of invocations, etc.
\item The state of the infrastructure: the location of nodes, the communication infrastructure, their online available resources, which functions are assigned to them, the execution logs, etc.
\item The data sources: the format of data, the location of data sources, their infrastructure, etc. 
\item The state of local caches: what they have in the caches, what policy for cache they use, what is the size of their cache, etc.
\end{enumerate}

Having all of the above information in an online manner incurs heavy overhead on the provider. On the other hand, having partial information may lead to imprecise decisions by the controller. This challenge deserves attention from both research and industrial communities.

\subsection{Security and Privacy}
Security is an indispensable concern in any computation service, be it serverless or not. Various security challenges are common between serverless and other cloud services. Precht et al. \cite{prechtl2020investigating} reviewed some of those challenges. Here, we survey the security issues that specifically threaten the normal operation of serverless service. We also consider the privacy of users in such environments.

\subsubsection{Authentication and authorization}
The foremost security challenge in any serverless scheme is how to authenticate applications so that only legitimate ones can use the available functions. Without authentication, a freeloader can use the available resources of the victims. A common approach to counter these attacks is the usage of authentication tokens in the header of requests. JWT is an example of such tokens \cite{jones2012json}. Amazon's Lambda currently implemented such a scheme that uses a bearer token authentication strategy to determine the caller's identity \cite{lambdaAuth}. However, if the request is sent through an unsecured channel, the attacker could simply extract the token and reuse it in another application. Using SSL/TLS protocols, this type of threat could be handled. However, there are cases in which these sophisticated public-key-based protocols are beyond the capabilities of the application's hardware. Very low-power IoT devices are an example of such hardware. Request signing is already proposed for such scenarios which incur lower resource consumption \cite{swedha2018analysis}. Albeit, we think the design and implementation of security protocols with minimal energy footprint for authentication in IoT and embedded devices is a promising direction for future research.

Serverless services are also susceptible to replay attacks. In these types of attacks, the attacker doesn't know about the content of the message being transmitted, however, they are interested in the effect of transmitting the message. So, the attacker captures the secured function execution request and replays it to sabotage the normal operation of the system. An example of such an attack is to replay a logout request indefinitely to prevent users from accessing their desired service. Detection and prevention approaches should be introduced for these types of attacks.


There is also the issue of authorization i.e. to specify which users or functions can invoke a certain function and to restrict others' access. This is different from the application-level authentication that we mentioned earlier. Here we authorize a function or a user to call another function. Lacking a proper authorization scheme can pose severe threats to the application's security. Recall that one of the envisioned advantages of serverless services, which we mentioned, is the ability to purchase off-the-shelf functions. Without authorization schemes, functions can be used without the consent of the application owner. To remedy this important issue Amazon Lambda uses a role-based access controls approach i.e. customers statically assign functions to roles that are associated with a set of permissions. The shortcoming of this approach is that it is restricted to a single function and the workflow-level access control is not considered. Sankaran et al. \cite{sankaran2020workflow} proposed WillIam that enables identity and access control for serverless workflows.

\subsubsection{Runtime Security}
In the wake of Meltdown \cite{lipp2018meltdown}, and Spectre \cite{kocher2018spectre} attacks, the vulnerability of applications against common execution environments has become one of the main security concerns. This issue is particularly severe in serverless environments since many functions from various owners are being executed in a shared execution environment. To counter these types of attacks, a lightweight and high-performance JavaScript engine is presented in \cite{brenner2019trust} that utilizes secure enclaves for the execution environments. The limitation of this work is that it suffers from high memory overhead and it only supports JavaScript. We think this is an area of research that deserves special attention from the research community. 

Runtime environments also can be customized to surveil the functions during execution to detect and prevent malicious activities. SecLambda \cite{jegan2020guarding} introduces a modified container runtime environment called runsec. It intercepts HTTP requests and I/O operations to check whether or not the function conforms with a set of predefined security policies. Trapeze \cite{alpernas2018secure} also adopts the same strategy i.e. it puts each serverless function in a sandbox that intercepts all interactions between the function and the rest of the world based on policy enforcement rules defined by their dynamic information flow control model. Valve  \cite{datta2020valve} proposes a serverless platform that performs runtime tracing and enforcement of information flow control. The downside of these approaches is that they have a fairly heavy impact on the performance of serverless services. More lightweight approaches need to be investigated for delay-sensitive functions.


\subsubsection{Resource Exhaustion Attacks}
The main focus of the attacker in these types of attacks is to over-utilize the resources of the victim to either disrupt the service or impose excessive financial/monetary loads. The victim in this type of attack can be both the service provider or the customer. An attacker may tamper with the application to send fraudulent requests to the provider. Although, the auto-scaling nature of serverless services can handle these situations, however, the load may go beyond the SLA with the provider and thus the provider may deny further requests, or at least it can impose a heavy financial load on the application owner. Monitoring approaches must be introduced for serverless providers to detect and mitigate these types of attacks.

Resource exhaustion attacks can also be established against the provider itself. These types of attacks would be particularly destructive for small to medium-sized providers. In this scenario, the attacker is familiar with the internal mechanisms of the provider or they can exploit it by studying the system's behavior. Using the knowledge, the attacker could launch a series of attacks that disrupt the normal operation of the system by intentionally preventing any optimization effort. For example, by knowing the packing strategy used by the provider, the attacker may inject fake dependencies to other data sources to prevent the function from being shipped near the data source which imposes heavy traffic inside the network of the provider.
 
\subsubsection{Privacy Issues}
There are many privacy-sensitive applications for serverless services especially in the area of IoT. For example, in a health-care application that gathers patients' data and then processes the data to infer certain conclusions, the privacy of the users is essential. The intent of the attacker here is not to alter the normal operation of the system as opposed to security attacks. Rather, they attempt to deduce knowledge about a user or a group of users, using a minimal set of gathered information. For example, in the domain of IoT healthcare system's an attacker may be interested in answering the question of whether a user has a heart condition or not.

There is much contextual data that an attacker could gather to infer knowledge about the victim. These are especially attainable when the network only uses application-layer security protocols. Here, we list some of these contextual data that can reveal sensitive information about the victim, along with their respective examples:

\begin{itemize}
\item Which function is invoked. For example in a serverless surveillance system, if the function \textit{gate-opened} is called. The attacker can deduce that someone has entered.
\item The sequence of function invocation, e.g., in a health-care monitoring system, a sequence of functions that are being invoked could reveal some kind of health condition in the patient.   
\item At which time/location a function is invoked. For example, an online retail store could reveal the vicinity in which a product is popular, which is interesting for the store's competitors.
\item The rate of function invocation. In our previous example i.e. surveillance system, this could reveal sensitive data about the traffic at the gates.
\end{itemize}

Comprehensive anonymity and disguise methods must be introduced for serverless services to prevent the breach of users' privacy.

\section{Conclusion}
In this paper, we surveyed some of the new opportunities that the vast adoption of serverless computing model will present. Then, we surveyed and categorized various serverless application domains. For each domain, we summarized the objectives for migrating to serverless services and assessed the aptness of the paradigm. We listed challenges that those applications faced and discussed existing solutions for them. We presented the areas that need further research investigations and identified open problems.

\bibliographystyle{ACM-Reference-Format}
\bibliography{shafiei}

\end{document}